\documentclass[a4paper,10pt]{article}
\usepackage{amssymb,amsmath}
\newcommand{\needstitle}[1]{\title{#1}}
\newcommand{\needsauthors}[1]{\author{#1}}
\newcommand{\needspresent}[1]{\underline{#1}}

\needstitle{Multidimensional the Ricci-flat  spaces \\[2mm]defined by
nonlinear equations}

\needsauthors{\needspresent{Valerii Dryuma}}

\begin{document}
\maketitle \vspace{-200pt} {\noindent\sl Conference Zakharov70",
\\Chernogolovka, August 1- 8, 2009.} \vspace{150pt}

\begin{itemize}
         \item{\small Institute of Mathematics and Computer Scince, MD-2028 Chisinau, Republic of Moldova.}
\end{itemize}

\section{The Ricci flat 8D metric and KP-equation}
   In the papers \cite{Dr0}-\cite{Dr1} was showed that the eight-dimensional metric in local coordinates $(x,y,z,t,P,Q,U,V)$
\begin{equation}\label{metric}
^{8}ds^2=2\left(-P H_{11}-P\frac{\partial F}{\partial t}-H_{12} Q-2\Gamma^3_{11} U+H_{22} V\right)dx^2+$$$$+4\left(H_{11} Q-H_{12}V\right)dx dy+4\left(-\frac{\partial F}{\partial z}V+\frac{\partial F}{\partial t} U\right)dx dz+2(H_{11}U-H_{31}V)dy^2+$$$$+2dx dP+2dydQ+2dz dU+2 dtdV,
\end{equation}
is  the Ricci-flat $R_{ik}=0$ if the conditions on the functions  $H_{ij}=H_{ij}(y,z,t)$ , $\Gamma^3_{11}(y,z,t)$ and $F(y,z,t)$
\begin{equation}\label{kon}\frac{\partial H_{12}}{\partial y}-\frac{\partial H_{22}}{\partial t}=0,\quad
-\frac{\partial H_{11}}{\partial y}+\frac{\partial H_{21}}{\partial t}=0,\quad
-\frac{\partial H_{11}}{\partial z}+\frac{\partial H_{31}}{\partial t}=0,\end{equation}
\begin{equation}\label{kon1}
\frac{\partial \Gamma^3_{11}}{\partial z}=2\left(\frac{\partial F}{\partial t}\right)^2+2H_{11}\frac{\partial F}{\partial t}+2(H_{11})^2.
\end{equation}
are hold.

    The metric (\ref{metric}) has the form
\begin{equation}\label{exten}
ds^2=-\Gamma^i_{jk}\xi_i dx^j dx^k+2d\xi_k dx^k \end{equation} and
it is an example of  Riemann extension of affinely-connected the
four dimensional space in local coordinates $x^l$ with
 symmetric connection $\Gamma^i_{jk}(x^l)=\Gamma^i_{kj}(x^l)$.
    From the system (\ref{kon})
after the substitution \cite{Kon1}
$$
H_{11}=-\frac{1}{2}u(y,z,t),\quad H_{12}=-\frac{1}{3}v(y,z,t)
,\quad H_{21}=-\frac{2}{3} v(y,z,t)-\frac{1}{2}\frac{\partial u(y,z,t)}{\partial t}
 $$
$$
H_{31}=-\frac{3}{4} w(y,z,t)+\frac{3}{8} u(y,z,t)^2-\frac{\partial v(y,z,t)}{\partial t}-\frac{1}{2}\frac{\partial^2 u(y,z,t)}{\partial t^2},
$$
$$
 H_{22}=-\frac{1}{2} w(y,z,t)+\frac{1}{2} u(y,z,t)^2-\frac{\partial v(y,z,t)}{\partial t}-\frac{1}{2}\frac{\partial^2 u(y,z,t)}{\partial t^2}+\frac{\partial u(y,z,t)}{\partial y},
$$
the famous KP-equation follows
\begin{equation}\label{KPD}
\frac{\partial}{\partial t}\left(\frac{\partial u(y,z,t)}{\partial z}-\frac{3}{2}u(y,z,t)\frac{\partial u(y,z,t)}{\partial t}
 -\frac{1}{4}\frac{\partial^3 u(y,z,t)}{\partial t^3}\right)=\frac{3}{4}\frac{\partial^2 u(y,z,t)}{\partial y^2}.
\end{equation}

\section{4D the Ricci-flat affinely connected subspace and KP-equation}
The main result of \cite{Dr1} is the following
\\[8pt]
\noindent \textbf{Theorem} The four-dimensional affinely connected space with
non zero coefficients of connection $\Gamma^i_{jk}=\Gamma^i_{kj}$ of the form
 $$
\Gamma^1_{11}=H_{11}+\frac{\partial F}{\partial t},\quad \Gamma^2_{11}=H_{12},\quad \Gamma^3_{11}=\Gamma^3_{11},,\quad \Gamma^2_{12}=H_{11},\quad
 \Gamma^3_{13}=-\frac{\partial F}{\partial t}
$$
$$
\Gamma^3_{22}=-H_{11},\quad \Gamma^4_{11}=-H_{22},\quad \Gamma^4_{12}=H_{21},\quad \Gamma^4_{13}=\frac{\partial F}{\partial z},\quad \Gamma^4_{22}=H_{31}
$$
is a Ricci flat
$$
R_{ij}=\partial_k\Gamma^k_{ij}-\partial_i\Gamma^k_{kj}+\Gamma^k_{kl}\Gamma^l_{ij}-\Gamma^k_{im}\Gamma^m_{kj}=0,
$$
if the conditions (\ref{kon}-\ref{kon1} ) hold.

  The problem of metrizability such type of connection is important
in theory of 3-dim manifolds now it is open question.

\section{6D  the Ricci-flat defined by KP-equation}

     As particular case we consider  the metrics (\ref{exten}) of the form
\begin{equation}\label{extKP}
^{6}{{\it ds}}^{2}=$$$$=4\,\left ({\frac {\partial }{\partial
x}}u(x,y,t)\right )P{\it dx}\,{ \it dt}+2\,{\it dx}\,{\it
dP}+4\,\left ({\frac {\partial }{\partial y} }u(x,y,t)\right
)P{\it dy}\,{\it dt}+2\,{\it dy}\,{\it dQ}+$$$$+\left (-2\,
Pu(x,y,t){\frac {\partial }{\partial x}}u(x,y,t)-2\,P{\frac
{\partial ^{3}}{\partial {x}^{3}}}u(x,y,t)-2\,\mu\,\left ({\frac
{\partial }{
\partial y}}u(x,y,t)\right )Q+2\,\left ({\frac {\partial }{\partial x}
}u(x,y,t)\right )U\right ){{\it dt}}^{2}+$$$$+2\,{\it dt}\,{\it
dU} .
\end{equation}

   The Ricci tensor such type of the metric
$$ R_{i j}=$$$$=\left [\begin {array}{cccccc} 0&0&0&0&0&0
\\\noalign{\medskip}0&0&0&0&0&0\\\noalign{\medskip}0&0&\left ({\frac {
\partial }{\partial x}}u(x,y,t)\right )^{2}\!+\!u(x,y,t){\frac {\partial ^
{2}}{\partial {x}^{2}}}u(x,y,t)\!+\!{\frac {\partial
^{4}}{\partial {x}^{4 }}}u(x,y,t)\!+\!{\frac {\partial
^{2}}{\partial t\partial x}}u(x,y,t)\!+\!\mu \,{\frac {\partial
^{2}}{\partial {y}^{2}}}u(x,y,t)&0&0&0
\\\noalign{\medskip}0&0&0&0&0&0\\\noalign{\medskip}0&0&0&0&0&0
\\\noalign{\medskip}0&0&0&0&0&0\end {array}\right ]
 ]
$$ has one component and it  is equal to zero on solutions of the
KP-equation
$$
 \frac{\partial( u_t+uu_x+u_{xxx})}{\partial
x}+\mu\frac{\partial^2 u }{\partial y^2}=0.
 $$

\noindent \textbf{Remark}

The metric (\ref{extKP}) arises from the  Riemann extension of the
3D Einstein-Weyl space
 $$
 ^{3}ds^2=dy^2-4dxdt-4u(x,y,t)dt^2,\quad \nu=-4u_xdt,
  $$
 associated with the DKP-equation
 $$
 \frac{\partial(u_t-uu_x)}{\partial x}=\frac{\partial^2 u}{\partial
 y^2},
 $$
which is obtained from the Einstein-Weyl conditions on the Ricci tensor
   $$
   R_{ij}+1/2\nabla(_i\nu_j)+1/4\nu_i\nu_j-1/3(R+1/2\nabla^k\nu_k+1/4\nu^k\nu_k)h_{ij}=0
   $$

\section{6D the Ricci-flat metrics associated with KdF-equation}

     3D-Riemann  metric of the form
\begin{equation}\label{extKdf}
{{\it ds}}^{2}={y}^{2}{{\it dx}}^{2}+ \left( l \left( x,z \right) {y}^
{2}-1/2 \right) {\it dx}\,{\it dz}+2\,{\it dy}\,{\it dz}+ $$$$+\left(
 \left( l \left( x,z \right)  \right) ^{2}{y}^{2}-2\, \left( {\frac {
\partial }{\partial x}}l \left( x,z \right)  \right) y+l \left( x
\cdot z \right)  \right) {{\it dz}}^{2}
\end{equation}
is a flat $R_{i j k l}=0$ if the function $l(x,z)$ satisfies the
 KdF-equation (Dryuma,2006)
\begin{equation}\label{Kdf}
{\frac {\partial ^{3}}{\partial {x}^{3}}}l \left( x,z \right) +{\frac
{\partial }{\partial z}}l \left( x,z \right) -3\,l \left( x,z \right)
{\frac {\partial }{\partial x}}l \left( x,z\right )=0.
\end{equation}

     It has $13$ Christoffel symbols.

    Some of them are in form
$$
\Gamma^1_{11}=
1/2\,{\frac {-1+2\,l \left( x,z \right) {y}^{2}}{y}},\quad \Gamma^1_{12}={y}^{-1},\quad
\Gamma^1_{13}=
1/2\,{\frac { \left( -1+2\,l \left( x,z \right) {y}^{2} \right) l
 \left( x,z \right) }{y}},$$$$
\Gamma^1_{23}=
{\frac {l \left( x,z \right) }{y}},\quad
$$
$$
\Gamma^1_{33}=1/2\,{\frac {2\, \left( {\frac {\partial }{\partial z}}l \left( x,z
 \right)  \right) y-4\,l \left( x,z \right) y{\frac {\partial }{
\partial x}}l \left( x,z \right) +2\,{\frac {\partial ^{2}}{\partial {
x}^{2}}}l \left( x,z \right) - \left( l \left( x,z \right)  \right) ^{
2}+2\, \left( l \left( x,z \right)  \right) ^{3}{y}^{2}}{y}},
$$
$$
\Gamma^2_{11}=
-1/4\,{\frac {4\,{y}^{3}{\frac {\partial }{\partial x}}l \left( x,z
 \right) -8\,l \left( x,z \right) {y}^{2}+1}{y}}.
$$
...
$$
\Gamma^3_{11}=-y,\quad \Gamma^3_{13}=-l \left( x,z \right) y,\quad \Gamma^3_{33}=- \left( l \left( x,z \right)  \right) ^{2}y+{\frac {\partial }{
\partial x}}l \left( x,z \right).
$$

     Six-dimensional Riemann extension of the metrics (\ref{extKdf}) is the space in local coordinates $(x,y,z,\xi_1,\xi_2,\xi_3$
 and it is defined by the expression
\begin{equation}\label{ext}
ds^2=-2\Gamma^k_{ij}\xi_k dx^i dx^j+2dx^k \xi_k
\end{equation}
with a given coefficients $\Gamma^k_{ij}$.

Such metric is a flat on solutions of the KdF-equation
(\ref{Kdf}).

    To obtain an examples of non a flat $R_{i j k l}\neq 0$ the six-dimensional metrics associated with the KdF-equation
it is necessary to introduce an additional terms into the
expression (\ref{ext}).

     As example the change of the component of metric
\[
g_{13}:=l(x,z)y^2-1/2
\]
on the following
\[
g_{1,3}:=l(x,z)y^2-1
\]
change radically  the  Ricci-tensor of the space.

From
$$
Matrix(3, 3,(1, 1) = 0, (1, 2) = 0, (1, 3) = 0, (2, 1) = 0, (2, 2) = 0, (2, 3) = 0,
$$
$$
(3, 1) = 0, (3, 2) = 0, (3, 3) = (-l_{xxx}-l_z+3ll_x)/y,
$$
on the
$$
Matrix(3, 3,(1, 1) = 0, (1, 2) = 0, (1, 3) = -(1/2)l_x/y, (2, 1) = 0, (2, 2) = 0, (2, 3) = 0, $$$$(3, 1) = -(1/2)l_x/y, (3, 2) = 0,\quad
 (3, 3) = (1/2)(-2l_{xxx}y^2-l_{xx}y-3l_z y^2+7ly^2l_x+l_x)/y^3.
$$

    In a six-dimensional case it is possible to change components of metric such a way
  that metric will be Ricci flat $R_{ik}=0$ on solutions of the KdF-equation
but not a flat $R_{ijkl}\neq 0$.

\section{6D Heavenly metric and Special Lagrangian equation}

     We study six-dimensional generalization of the Heavenly
     metrics
\begin{equation}\label{dryuma:eq3}
{{\it ds}}^{2}={\it dx}\,{\it du}+{\it dy}\,{\it dv}+{\it
dz}\,{\it dw }+A(x,y,z,u,v,w){{\it
du}}^{2}+$$$$+2\,B(x,y,z,u,v,w){\it du}\,{\it dv}+2\,
E(x,y,z,u,v,w){\it du}\,{\it dw}+C(x,y,z,u,v,w){{\it
dv}}^{2}+$$$$+2\,H(x,y ,z,u,v,w){\it dv}\,{\it
dw}+F(x,y,z,u,v,w){{\it dw}}^{2}.
\end{equation}

     The Ricci tensor of the metric ~(\ref{dryuma:eq3}) has a fifteen components.

     Nine of them  are equal to zero due the conditions
\begin{equation}\label{dryuma:eq4}
   {\frac {\partial }{\partial u}}E(x,y,z,u,v,w)+{\frac {\partial }{
\partial v}}H(x,y,z,u,v,w)+{\frac {\partial }{\partial w}}F(x,y,z,u,v,
w)=0,$$$${\frac {\partial }{\partial u}}B(x,y,z,u,v,w)+{\frac
{\partial }{
\partial v}}C(x,y,z,u,v,w)+{\frac {\partial }{\partial w}}H(x,y,z,u,v,
w)=0,$$$${\frac {\partial }{\partial u}}A(x,y,z,u,v,w)+{\frac
{\partial }{
\partial v}}B(x,y,z,u,v,w)+{\frac {\partial }{\partial w}}E(x,y,z,u,v.
w)=0.
\end{equation}

   This system of equation has a solutions  depending  from an arbitrary functions.

   In  a simplest case we have the solution
\[
A(x,y,z,u,v,w)=\left ({\frac {\partial ^{2}}{\partial
{z}^{2}}}f(x,y,z ,u,v,w)\right ){\frac {\partial ^{2}}{\partial
{y}^{2}}}f(x,y,z,u,v,w) -\left ({\frac {\partial ^{2}}{\partial
y\partial z}}f(x,y,z,u,v,w) \right )^{2},\]\[ C(x,y,z,u,v,w)=\left
({\frac {\partial ^{2}}{\partial {x}^{2}}}f(x,y,z ,u,v,w)\right
){\frac {\partial ^{2}}{\partial {z}^{2}}}f(x,y,z,u,v,w) -\left
({\frac {\partial ^{2}}{\partial x\partial z}}f(x,y,z,u,v,w)
\right )^{2},\]\[ F(x,y,z,u,v,w)=\left ({\frac {\partial
^{2}}{\partial {x}^{2}}}f(x,y,z ,u,v,w)\right ){\frac {\partial
^{2}}{\partial {y}^{2}}}f(x,y,z,u,v,w) -\left ({\frac {\partial
^{2}}{\partial x\partial y}}f(x,y,z,u,v,w) \right )^{2},\]\[\quad
E(x,y,z,u,v,w)=\left ({\frac {\partial ^{2}}{\partial y\partial
z}}f(x ,y,z,u,v,w)\right ){\frac {\partial ^{2}}{\partial
x\partial y}}f(x,y, z,u,v,w)-\]\[-\left ({\frac {\partial
^{2}}{\partial x\partial z}}f(x,y,z,u ,v,w)\right ){\frac
{\partial ^{2}}{\partial {y}^{2}}}f(x,y,z,u,v,w)
\]\[B(x,y,z,u,v,w)=\left ({\frac {\partial ^{2}}{\partial
x\partial z}}f(x ,y,z,u,v,w)\right ){\frac {\partial
^{2}}{\partial y\partial z}}f(x,y, z,u,v,w)-\]\[-\left ({\frac
{\partial ^{2}}{\partial x\partial y}}f(x,y,z,u ,v,w)\right
){\frac {\partial ^{2}}{\partial {z}^{2}}}f(x,y,z,u,v,w)\]\[
H(x,y,z,u,v,w)=\left ({\frac {\partial ^{2}}{\partial x\partial
z}}f(x ,y,z,u,v,w)\right ){\frac {\partial ^{2}}{\partial
x\partial y}}f(x,y, z,u,v,w)-\]\[-\left ({\frac {\partial
^{2}}{\partial y\partial z}}f(x,y,z,u ,v,w)\right ){\frac
{\partial ^{2}}{\partial {x}^{2}}}f(x,y,z,u,v,w)
\]
depending from one arbitrary function.

      At this conditions the  six-dimensional metric looks as
\begin{equation}\label{dryuma:eq5}
^{6}ds^2=\left ({\frac {\partial ^{2}}{\partial {w}^{2}}}K(\vec x)
{\frac {\partial ^{2}}{\partial {v}^{2}}}K(\vec x)-\left ({\frac
{\partial ^{2}}{\partial v\partial w}}K(\vec x)\right )^{ 2}\right
){d{{x}}}^{2}\!+\!$$$$+2\,\left ({\frac {\partial ^{2}}{
\partial u\partial w}}K(\vec x){\frac {\partial ^{2}}{
\partial v\partial w}}K(\vec x)\!-\!{\frac {\partial ^{2}}{
\partial {w}^{2}}}K(\vec x){\frac {\partial ^{2}}{
\partial u\partial v}}K(\vec x)\right )d{{x}}d{{y}}\!+\!$$$$+\!\left (
{\frac {\partial ^{2}}{\partial {u}^{2}}}K(\vec x) {\frac
{\partial ^{2}}{\partial {w}^{2}}}K(\vec x)\!-\!\left ({\frac
{\partial ^{2}}{\partial u\partial w}}K(\vec x)\right )^{2} \right
){d{{y}}}^{2}\!+\!$$$$+2\,\left ({\frac {\partial ^{2}}{
\partial v\partial w}}K(\vec x){\frac {\partial ^{2}}{
\partial u\partial v}}K(\vec x)\!-\!{\frac {\partial ^{2}}{
\partial u\partial w}}K(\vec x){\frac {\partial ^{2}}{
\partial {v}^{2}}}K(\vec x)\right )d{{x}}d{{z}}\!+\!$$$$\!+\!\left ({
\frac {\partial ^{2}}{\partial {v}^{2}}}K(\vec x){\frac {
\partial ^{2}}{\partial {u}^{2}}}K(\vec x)\!-\!\left ({\frac {
\partial ^{2}}{\partial u\partial v}}K(\vec x)\right )^{2}\right
){d{{z}}}^{2}\!+\!$$$$\!+\!2\,\left ({\frac {\partial
^{2}}{\partial u
\partial w}}K(\vec x){\frac {\partial ^{2}}{\partial u
\partial v}}K(\vec x)\!-\!{\frac {\partial ^{2}}{\partial {u}^
{2}}}K(\vec x){\frac {\partial ^{2}}{\partial v\partial w }}K(\vec
x)\right )d{{z}}d{{y}}\!+\!$$$$\!+\!d{{x}}d{{u}}+d{{y}}d{{v}}+d
{{z}}d{{w}}
\end{equation}
where $K(\vec x)=K(x,y,z,u,v,w)$ is arbitrary function.

   The Ricci tensor $R_{ij}$ of the metric ~(\ref{dryuma:eq5}) has a six components.

   All equations
   \[
   R_{ij}=0
   \]
   after the substitution
   \[
   K(x,y,z,u,v,w)=\phi(y+v+x,z+w+x)
\]
 are reduced to the one equation
\begin{equation}\label{dryuma:eq6}
-\left ({\frac {\partial ^{4}}{\partial \xi\partial {\rho}^{2}
\partial \xi}}\phi(\xi,\rho)\right ){\frac {\partial ^{2}}{\partial {
\xi}^{2}}}\phi(\xi,\rho)+2\,\left ({\frac {\partial ^{3}}{\partial
\xi
\partial \rho\partial \xi}}\phi(\xi,\rho)\right )^{2}-2\,\left ({
\frac {\partial ^{3}}{\partial {\rho}^{2}\partial
\xi}}\phi(\xi,\rho) \right ){\frac {\partial ^{3}}{\partial
{\xi}^{3}}}\phi(\xi,\rho)-$$$$- \left ({\frac {\partial
^{2}}{\partial {\rho}^{2}}}\phi(\xi,\rho) \right ){\frac {\partial
^{4}}{\partial {\xi}^{4}}}\phi(\xi,\rho)+2\, \left ({\frac
{\partial ^{2}}{\partial \rho\partial \xi}}\phi(\xi,\rho )\right
){\frac {\partial ^{4}}{\partial {\xi}^{2}\partial \rho
\partial \xi}}\phi(\xi,\rho)+$$$$+2\,\left ({\frac {\partial ^{3}}{
\partial {\rho}^{2}\partial \xi}}\phi(\xi,\rho)\right )^{2}-\left ({
\frac {\partial ^{4}}{\partial {\rho}^{4}}}\phi(\xi,\rho)\right ){
\frac {\partial ^{2}}{\partial {\xi}^{2}}}\phi(\xi,\rho)-2\,\left
({ \frac {\partial ^{3}}{\partial {\rho}^{3}}}\phi(\xi,\rho)\right
){ \frac {\partial ^{3}}{\partial \xi\partial \rho\partial
\xi}}\phi(\xi, \rho)-$$$$-\left ({\frac {\partial ^{2}}{\partial
{\rho}^{2}}}\phi(\xi,\rho )\right ){\frac {\partial ^{4}}{\partial
\xi\partial {\rho}^{2}
\partial \xi}}\phi(\xi,\rho)+2\,\left ({\frac {\partial ^{2}}{
\partial \rho\partial \xi}}\phi(\xi,\rho)\right ){\frac {\partial ^{4}
}{\partial {\rho}^{3}\partial \xi}}\phi(\xi,\rho)=0,
\end{equation}
 where
\[
\xi=x+y+v,\quad \rho=z+x+w.
\]

    In compact form this equation can be rewritten as
    \[
    \Delta\psi(\xi,\rho)=0
    \]
where
\[
\psi(\xi,~\rho)=\left ({\frac {\partial ^{2}}{\partial
{\xi}^{2}}}\phi(\xi,\rho) \right ){\frac {\partial ^{2}}{\partial
{\rho}^{2}}}\phi(\xi,\rho)- \left ({\frac {\partial ^{2}}{\partial
\xi\partial \rho}}\phi(\xi,\rho )\right )^{2}
\]
and
\[
\Delta=\frac{\partial^2}{ \partial \xi^2}+\frac{\partial^2}{
\partial \rho^2}
\]
is the Laplace operator.

     Its solutions give the Ricci-flat examples of the metric
     ~(\ref{dryuma:eq5}).

\subsection{ The Beltrami parameters}

    To the investigation of the properties of the metrics ~(\ref{dryuma:eq5}) can be considered
    two invariant equations defined by the first
\[
\Delta \psi= g^{i j}\frac{\partial \psi}{ \partial
x^i}\frac{\partial \psi}{
\partial x^j}
\]
 and the second
\[
\Box \psi=g^{ij}\left(\frac{\partial^2}{ \partial x^i
\partial x^j}-\Gamma^k_{ij}\frac{\partial}{
\partial x^k}\right)\psi
\]
Beltrami parameters.

    To the metric ~(\ref{dryuma:eq5}) the equation $\Box \phi=0$ looks
    as ($K=\phi(\vec x)$)
\begin{equation}\label{dryuma:eq7}
 {\frac {\partial ^{2}}{\partial u\partial x}}\phi(\vec x)+{
\frac {\partial ^{2}}{\partial w\partial z}}\phi(\vec x)+{ \frac
{\partial ^{2}}{\partial v\partial y}}\phi(\vec x)- \left ({\frac
{\partial ^{2}}{\partial {y}^{2}}}\phi(\vec x) \right )\left
({\frac {\partial ^{2}}{\partial {x}^{2}}}f(\vec x) \right ){\frac
{\partial ^{2}}{\partial {z}^{2}}}f(\vec x)+$$$$+2\, \left ({\frac
{\partial ^{2}}{\partial x\partial z}}\phi(\vec x) \right )\left
({\frac {\partial ^{2}}{\partial x\partial z}}f(\vec x)\right
){\frac {\partial ^{2}}{\partial {y}^{2}}}f(\vec x)-2 \,\left
({\frac {\partial ^{2}}{\partial x\partial z}}\phi(\vec x)\right
)\left ({\frac {\partial ^{2}}{\partial y\partial z}}f(\vec
x)\right ){\frac {\partial ^{2}}{\partial x\partial y}}f(\vec
x)+$$$$+2\,\left ({\frac {\partial ^{2}}{\partial x\partial
y}}\phi(\vec x)\right )\left ({\frac {\partial ^{2}}{\partial
x\partial y}}f(\vec x)\right ){\frac {\partial ^{2}}{\partial
{z}^{2}}}f(\vec x)-\left ({\frac {\partial ^{2}}{\partial
{x}^{2}}}\phi(\vec x )\right )\left ({\frac {\partial
^{2}}{\partial {z}^{2}}}f(\vec x )\right ){\frac {\partial
^{2}}{\partial {y}^{2}}}f(\vec x)+$$$$+2\, \left ({\frac {\partial
^{2}}{\partial y\partial z}}\phi(\vec x) \right )\left ({\frac
{\partial ^{2}}{\partial y\partial z}}f(\vec x)\right ){\frac
{\partial ^{2}}{\partial {x}^{2}}}f(\vec x)-2 \,\left ({\frac
{\partial ^{2}}{\partial y\partial z}}\phi(\vec x )\right )\left
({\frac {\partial ^{2}}{\partial x\partial z}}f(\vec x)\right
){\frac {\partial ^{2}}{\partial x\partial y}}f(\vec x)-$$$$-\left
({\frac {\partial ^{2}}{\partial {z}^{2}}}\phi(\vec x)\right
)\left ({\frac {\partial ^{2}}{\partial {x}^{2}}}f(\vec x)\right
){\frac {\partial ^{2}}{\partial {y}^{2}}}f(\vec x)-2\, \left
({\frac {\partial ^{2}}{\partial x\partial y}}\phi(\vec x) \right
)\left ({\frac {\partial ^{2}}{\partial x\partial z}}f(\vec
x)\right ){\frac {\partial ^{2}}{\partial y\partial z}}f(\vec
x)+$$$$+\left ({\frac {\partial ^{2}}{\partial {z}^{2}}}\phi(\vec
x) \right )\left ({\frac {\partial ^{2}}{\partial x\partial
y}}f(\vec x)\right )^{2}+\left ({\frac {\partial ^{2}}{\partial
{y}^{2}}} \phi(\vec x)\right )\left ({\frac {\partial
^{2}}{\partial x
\partial z}}f(\vec x)\right )^{2}+$$$$+\left ({\frac {\partial ^{2}
}{\partial {x}^{2}}}\phi(\vec x)\right )\left ({\frac {\partial ^
{2}}{\partial y\partial z}}f(\vec x)\right )^{2}=0.
\end{equation}

   In particular case the equation ~(\ref{dryuma:eq7}) after the substitution
   \[
   \phi(\vec x)=f(\vec x)
   \]
takes the form
\begin{equation}\label{dryuma:eq8}
{\frac {\partial ^{2}}{\partial u\partial x}}f(\vec x)+{ \frac
{\partial ^{2}}{\partial w\partial z}}f(\vec x)+{\frac {
\partial ^{2}}{\partial v\partial y}}f(\vec x)-3\,\left ({\frac
{\partial ^{2}}{\partial {y}^{2}}}f(\vec x)\right )\left ({\frac
{\partial ^{2}}{\partial {x}^{2}}}f(\vec x)\right ){\frac {
\partial ^{2}}{\partial {z}^{2}}}f(\vec x)+$$$$+3\,\left ({\frac {
\partial ^{2}}{\partial x\partial z}}f(\vec x)\right )^{2}{\frac
{\partial ^{2}}{\partial {y}^{2}}}f(\vec x)-6\,\left ({\frac {
\partial ^{2}}{\partial x\partial z}}f(\vec x)\right )\left ({
\frac {\partial ^{2}}{\partial y\partial z}}f(\vec x)\right ){
\frac {\partial ^{2}}{\partial x\partial y}}f(\vec
x)+$$$$+3\,\left ( {\frac {\partial ^{2}}{\partial x\partial
y}}f(\vec x)\right )^{2 }{\frac {\partial ^{2}}{\partial
{z}^{2}}}f(\vec x)+3\,\left ({ \frac {\partial ^{2}}{\partial
y\partial z}}f(\vec x)\right )^{2} {\frac {\partial ^{2}}{\partial
{x}^{2}}}f(\vec x)=0. \end{equation}

   After the change of variables
\[
f(\vec x)=f(x,y,z,u,v,w)=h(x+u,v+y,w+z)=h(\eta,\xi,\rho)
\]
the equation ~(\ref{dryuma:eq8}) is reduced to the form
\begin{equation}\label{dryuma:eq9}
 {\frac {\partial
^{2}}{\partial {\eta}^{2}}}h(\eta,\xi,\rho)+{\frac {
\partial ^{2}}{\partial {\rho}^{2}}}h(\eta,\xi,\rho)+{\frac {\partial
^{2}}{\partial {\xi}^{2}}}h(\eta,\xi,\rho)+3\,\left ({\frac
{\partial ^{2}}{\partial \eta\partial \rho}}h(\eta,\xi,\rho)\right
)^{2}{\frac {
\partial ^{2}}{\partial {\xi}^{2}}}h(\eta,\xi,\rho)-$$$$-6\,\left ({\frac {
\partial ^{2}}{\partial \eta\partial \rho}}h(\eta,\xi,\rho)\right )
\left ({\frac {\partial ^{2}}{\partial \rho\partial
\xi}}h(\eta,\xi, \rho)\right ){\frac {\partial ^{2}}{\partial
\eta\partial \xi}}h(\eta, \xi,\rho)+3\,\left ({\frac {\partial
^{2}}{\partial \eta\partial \xi}} h(\eta,\xi,\rho)\right
)^{2}{\frac {\partial ^{2}}{\partial {\rho}^{2}
}}h(\eta,\xi,\rho)-$$$$-3\,\left ({\frac {\partial ^{2}}{\partial
{\xi}^{2} }}h(\eta,\xi,\rho)\right )\left ({\frac {\partial
^{2}}{\partial {\eta }^{2}}}h(\eta,\xi,\rho)\right ){\frac
{\partial ^{2}}{\partial {\rho}^ {2}}}h(\eta,\xi,\rho)+3\,\left
({\frac {\partial ^{2}}{\partial \rho
\partial \xi}}h(\eta,\xi,\rho)\right )^{2}{\frac {\partial ^{2}}{
\partial {\eta}^{2}}}h(\eta,\xi,\rho)=0
 \end{equation}
or
\[
\Delta h(\eta,\xi,\rho)-3\det\left [\begin {array}{ccc} {\frac
{\partial ^{2}}{
\partial {\eta}^{2}}}h(\eta,\xi,\rho)&{\frac {\partial ^{2}}{
\partial \eta\partial \xi}}h(\eta,\xi,\rho)&{\frac {\partial ^{2}}
{\partial \eta\partial \rho}}h(\eta,\xi,\rho)\\\noalign{\medskip}{
\frac {\partial ^{2}}{\partial \eta\partial \xi}}h(\eta,\xi,\rho)&
{\frac {\partial ^{2}}{\partial {\xi}^{2}}}h(\eta,\xi,\rho)&{
\frac {\partial ^{2}}{\partial \rho\partial \xi}}h(\eta,\xi,\rho)
\\\noalign{\medskip}{\frac {\partial ^{2}}{\partial \eta\partial
\rho}}h(\eta,\xi,\rho)&{\frac {\partial ^{2}}{\partial \rho
\partial \xi}}h(\eta,\xi,\rho)&{\frac {\partial ^{2}}{\partial {
\rho}^{2}}}h(\eta,\xi,\rho)\end {array}\right ]=0,
\]
where
\[ \Delta=\frac{\partial^2}{ \partial
\eta^2}+\frac{\partial^2}{ \partial \xi^2}+\frac{\partial^2}{
\partial \rho^2}
\]
is a three-dimensional Laplace operator.

   The equation ~(\ref{dryuma:eq9}) is a famous Harvey-Lawson "Special Lagrangian" equation
   having an important applications in theory of Calabi-Yau manifolds and mirror symmetry.

\subsection{The simplest solutions}

          To obtain  particular solutions of the partial nonlinear
       differential equation
\begin{equation}\label{Dr3}
F(x,y,z,f_x,f_y,f_z,f_{xx},f_{xy},f_{xz},f_{yy},f_{yz},f_{xxx},f_{xyy},f_{xxy},..)=0
 \end{equation}
    can be applied a following approach.

      We use the following parametric presentation of the functions and variables
\begin{equation}\label{Dr4}
f(x,y,z)\rightarrow u(x,t,z),\quad y \rightarrow v(x,t,z),\quad
f_x\rightarrow u_x-\frac{u_t}{v_t}v_x,$$$$ f_z\rightarrow
u_z-\frac{u_t}{v_t}v_z,\quad f_y \rightarrow \frac{u_t}{v_t},
\quad f_{yy} \rightarrow \frac{(\frac{u_t}{v_t})_t}{v_t}, \quad
f_{xy} \rightarrow \frac{(u_x-\frac{u_t}{v_t}v_x)_t}{v_t},...
\end{equation}
where variable $t$ is considered as parameter.

  Remark that conditions of the type
   \[
   f_{xy}=f_{yx},\quad f_{xz}=f_{zx}...
   \]
hold at the such type of presentation.

  In result instead of equation (\ref{Dr3}) one get the
  relation between the new variables $u(x,t,z)$ and $v(x,t,z)$ and
  their partial derivatives
\begin{equation}\label{Dr5}
\Psi(u,v,u_x,u_z,u_t,v_x,v_z,v_t,...)=0.
  \end{equation}

    This relation coincides with initial p.d.e at the condition $v(x,t,z)=t$
    and takes more general form after presentation of the functions $u,v$ in form $u(x,t,z)=F(\omega,\omega_t...)$
    and $v(x,t,z,s)=\Phi(\omega,\omega_t...)$ with some function $\omega(x,t,z)$  .

      In result of change variables and function with accordance of (\ref{Dr4}) the equation
    ~(\ref{dryuma:eq9})
\[
{\frac {\partial ^{2}}{\partial {x}^{2}}}h(x,y,z)+{\frac {\partial
^{2 }}{\partial {z}^{2}}}h(x,y,z)+{\frac {\partial ^{2}}{\partial
{y}^{2}} }h(x,y,z)+3\,\left ({\frac {\partial ^{2}}{\partial
x\partial z}}h(x,y ,z)\right )^{2}{\frac {\partial ^{2}}{\partial
{y}^{2}}}h(x,y,z)-\]\[-6\, \left ({\frac {\partial ^{2}}{\partial
x\partial z}}h(x,y,z)\right ) \left ({\frac {\partial
^{2}}{\partial y\partial z}}h(x,y,z)\right ){ \frac {\partial
^{2}}{\partial x\partial y}}h(x,y,z)+3\,\left ({\frac {\partial
^{2}}{\partial x\partial y}}h(x,y,z)\right )^{2}{\frac {
\partial ^{2}}{\partial {z}^{2}}}h(x,y,z)-\]\[-3\,\left ({\frac {\partial ^
{2}}{\partial {y}^{2}}}h(x,y,z)\right )\left ({\frac {\partial
^{2}}{
\partial {x}^{2}}}h(x,y,z)\right ){\frac {\partial ^{2}}{\partial {z}^
{2}}}h(x,y,z)+3\,\left ({\frac {\partial ^{2}}{\partial y\partial
z}}h (x,y,z)\right )^{2}{\frac {\partial ^{2}}{\partial
{x}^{2}}}h(x,y,z)=0
\]
is transformed into  the relation (\ref{Dr5}).

This relation  after the substitution
\begin{equation}\label{Dr81}
u(x,t,z)=t{\frac {\partial }{\partial
t}}\omega(x,t,z)-\omega(x,t,z),\quad
 v(x,t,z)={\frac {\partial
}{\partial t}}\omega(x,t,z). \end{equation}
 takes the form of
p.d.e.
\begin{equation}\label{Dr8}
-3\,\left ({\frac {\partial ^{2}}{\partial {x}^{2}}}\omega(x,t,z)
\right ){\frac {\partial ^{2}}{\partial {z}^{2}}}\omega(x,t,z)+3\,
\left ({\frac {\partial ^{2}}{\partial x\partial z}}\omega(x,t,z)
\right )^{2}-\left ({\frac {\partial ^{2}}{\partial
{t}^{2}}}\omega(x, t,z)\right ){\frac {\partial ^{2}}{\partial
{x}^{2}}}\omega(x,t,z)+$$$$+1+ \left ({\frac {\partial
^{2}}{\partial t\partial z}}\omega(x,t,z) \right )^{2}+\left
({\frac {\partial ^{2}}{\partial t\partial x}} \omega(x,t,z)\right
)^{2}-\left ({\frac {\partial ^{2}}{\partial {t}^{
2}}}\omega(x,t,z)\right ){\frac {\partial ^{2}}{\partial {z}^{2}}}
\omega(x,t,z)=0.
\end{equation}

       The equation (\ref{Dr8}) consists from the 2D M-A equations
       with respect the variables $(x,t)$, $(x,z)$ and $(t,z)$.

        Let us consider some examples of its solutions.

       1. To the equation
\begin{equation}\label{Dr9}
\mu\,\Delta \left( f \right) +{\it Hess} \left( f \right) =0
\end{equation}

The substitution into (\ref{Dr9})
\[
\omega \left( x,t,z \right) =A \left( {x}^{2}+{z}^{2},t \right)
\]
lead to the equation on the function $A(\eta,t)$, ( $\eta=x^2+z^2$)
\[
-4\,\mu\, \left( {\frac {\partial ^{2}}{\partial \eta\partial t}}A
 \left( \eta,t \right)  \right) ^{2}\eta-8\, \left( {\frac {\partial ^
{2}}{\partial {\eta}^{2}}}A \left( \eta,t \right)  \right) \eta\,{
\frac {\partial }{\partial \eta}}A \left( \eta,t \right) -4\, \left( {
\frac {\partial }{\partial \eta}}A \left( \eta,t \right)  \right) ^{2}
+\]\[+4\,\mu\, \left( {\frac {\partial ^{2}}{\partial {t}^{2}}}A \left(
\eta,t \right)  \right) {\frac {\partial }{\partial \eta}}A \left(
\eta,t \right) -\mu+4\,\mu\, \left( {\frac {\partial ^{2}}{\partial {t
}^{2}}}A \left( \eta,t \right)  \right) {\frac {\partial ^{2}}{
\partial {\eta}^{2}}}A \left( \eta,t \right) =0.
\]

Its particular solution is
\[
A \left( \eta,t \right) =B \left( t \right) +\eta\,{e^{kt}},\]\[
B \left( t \right) ={\frac {{e^{kt}}}{\mu\,{k}^{2}}}+1/4\,{\frac {{e^{
-kt}}}{{k}^{2}}}+{\it \_C1}\,t+{\it \_C2}.
\]

    Now elimination of the parameter $t$ from the system
\[
y-{\frac {\partial }{\partial t}}\omega \left( x,t,z \right)=0,\quad
f \left( x,y,z \right) -t{\frac {\partial }{\partial t}}\omega \left(
x,t,z \right) +\omega \left( x,t,z \right)=0
\]
lead to the solution of the equation (\ref{Dr9})
\[
f \left( x,y,z \right) =- \left( 1-\ln  \left( {\frac {y\mu+\sqrt {T}}
{1+\mu\,{x}^{2}+\mu\,{z}^{2}}} \right) y\sqrt {T}+\ln  \left( 2
 \right) {y}^{2}\mu-\ln  \left( {\frac {y\mu+\sqrt {T}}{1+\mu\,{x}^{2}
+\mu\,{z}^{2}}} \right) {y}^{2}\mu\right)\left( y
\mu+\sqrt {T} \right) ^{-1}+\]\[+\left({y}^{2}\mu+\mu\,{x}^{2}\right)+\left(\ln
 \left( 2 \right) y\sqrt {T}+\mu\,{z}^{2}+y\sqrt {T} \right)  \left( y
\mu+\sqrt {T} \right) ^{-1},
\]
where
\[
\mu\, \left( {y}^{2}\mu+1+\mu\,{x}^{2}+\mu\,{z}^{2} \right) =T
\]

     2.  The substitution
\[
\omega \left( x,t,z \right) =A \left( x+t,z \right)
\]
into  (\ref{Dr9})
lead to the equation on the function $A(eta,z$ ($\eta=x+z$)
\[- \left( {\frac {\partial ^{2}}{\partial {\eta}^{2}}}A \left( \eta,z
 \right)  \right) {\frac {\partial ^{2}}{\partial {z}^{2}}}A \left(
\eta,z \right) -\mu\, \left( {\frac {\partial ^{2}}{\partial \eta
\partial z}}A \left( \eta,z \right)  \right) ^{2}+\mu\, \left( {\frac
{\partial ^{2}}{\partial {\eta}^{2}}}A \left( \eta,z \right)  \right)
{\frac {\partial ^{2}}{\partial {z}^{2}}}A \left( \eta,z \right) -\]\[-\mu+
 \left( {\frac {\partial ^{2}}{\partial \eta\partial z}}A \left( \eta,
z \right)  \right) ^{2}
 \]

    At the $\mu=-1/3$ we get the M-A equation
\begin{equation}\label{eq10}
- \left( {\frac {\partial ^{2}}{\partial {\eta}^{2}}}A \left( \eta,z
 \right)  \right) {\frac {\partial ^{2}}{\partial {z}^{2}}}A \left(
\eta,z \right) + \left( {\frac {\partial ^{2}}{\partial \eta\partial z
}}A \left( \eta,z \right)  \right) ^{2}+1/4=0
\end{equation}
which after the $(u,v)$-transformation is reduced to the Laplace equation
\[
4\,{\frac {\partial ^{2}}{\partial {\eta}^{2}}}\theta \left( \eta,\xi
 \right) +{\frac {\partial ^{2}}{\partial {\xi}^{2}}}\theta \left(
\eta,\xi \right) =0.
\]

     Its general solution
\[
\theta(\eta,\xi)=M(\eta+2 I \xi)+N(\eta-2 I \xi)
\]
allow us to construct particular solutions of the equation (\ref{eq10}) and
corresponding ``Special lagrangian equaion'' .

\section{Heavenly metrics and Yang-Mills equation}

    We consider following  generalization of Heavenly metrics
\begin{equation}\label{metr}
{{\it ds}}^{2}={\it dx}\,{\it du}+{\it dy}\,{\it dv}+{\it
dz}\,{\it dw }+A(x,y,z,p,q){{\it du}}^{2}+2\,B(x,y,z,p,q){\it
du}\,{\it dv}+$$$$+2\,E(x, y,z,p,q){\it du}\,{\it
dw}+C(x,y,z,p,q){{\it dv}}^{2}+2\,H(x,y,z,p,q){ \it dv}\,{\it
dw}+$$$$+F(x,y,z,p,q){{\it dw}}^{2}+{\it dp}\,{\it dq},
\end{equation}
 having a following components
     \[
     A(x,y,z,p,q)=\left ({\frac {\partial ^{2}}{\partial {z}^{2}}}f(x,y,z,p
,q)\right ){\frac {\partial ^{2}}{\partial {y}^{2}}}f(x,y,z,p,q)-
\left ({\frac {\partial ^{2}}{\partial y\partial z}}f(x,y,z,p,q)
\right )^{2},\]\[ C(x,y,z,p,q)=\left ({\frac {\partial
^{2}}{\partial {x}^{2}}}f(x,y,z,p ,q)\right ){\frac {\partial
^{2}}{\partial {z}^{2}}}f(x,y,z,p,q)- \left ({\frac {\partial
^{2}}{\partial x\partial z}}f(x,y,z,p,q) \right )^{2},\]\[
F(x,y,z,p,q)=\left ({\frac {\partial ^{2}}{\partial
{x}^{2}}}f(x,y,z,p ,q)\right ){\frac {\partial ^{2}}{\partial
{y}^{2}}}f(x,y,z,p,q)- \left ({\frac {\partial ^{2}}{\partial
x\partial y}}f(x,y,z,p,q) \right )^{2},\]\[ E(x,y,z,p,q)=\left
({\frac {\partial ^{2}}{\partial y\partial z}}f(x,y ,z,p,q)\right
){\frac {\partial ^{2}}{\partial x\partial y}}f(x,y,z,p, q)-\left
({\frac {\partial ^{2}}{\partial x\partial z}}f(x,y,z,p,q) \right
){\frac {\partial ^{2}}{\partial {y}^{2}}}f(x,y,z,p,q),\]\[
B(x,y,z,p,q)=\left ({\frac {\partial ^{2}}{\partial x\partial
z}}f(x,y ,z,p,q)\right ){\frac {\partial ^{2}}{\partial y\partial
z}}f(x,y,z,p, q)-\left ({\frac {\partial ^{2}}{\partial x\partial
y}}f(x,y,z,p,q) \right ){\frac {\partial ^{2}}{\partial
{z}^{2}}}f(x,y,z,p,q),\]\[ H(x,y,z,p,q)=\left ({\frac {\partial
^{2}}{\partial x\partial z}}f(x,y ,z,p,q)\right ){\frac {\partial
^{2}}{\partial x\partial y}}f(x,y,z,p, q)-\left ({\frac {\partial
^{2}}{\partial y\partial z}}f(x,y,z,p,q) \right ){\frac {\partial
^{2}}{\partial {x}^{2}}}f(x,y,z,p,q).
\]

    At these conditions from $21$ components of the  Ricci-tensor of the metric (~\ref{metr})
     only six components
\[
R_{uu}\neq 0,\quad R_{uv}\neq 0,\quad R_{uw}\neq 0,\quad
R_{vv}\neq 0,\quad R_{vw}\neq 0,\quad R_{ww}\neq 0.
\]
are different from zero.

  The equation
  \[
g^{ij}\left (\frac{\partial^2 \Psi}{\partial x^i \partial
x^j}-\Gamma^k_{ij}\frac{\partial \Psi}{\partial x^k}\right)=0,
\]
defined by the Laplace-Beltrami operator of the metric
(~\ref{metr}) depends on the function $f(x,y,z,p,q)$ and after the
substitution $f=\Psi$ takes the form
\[
{\frac {\partial ^{2}}{\partial p\partial
q}}f(x,y,z,p,q)+\]\[+3\,\left ({ \frac {\partial ^{2}}{\partial
x\partial y}}f(x,y,z,p,q)\right )^{2}{ \frac {\partial
^{2}}{\partial {z}^{2}}}f(x,y,z,p,q)-\]\[-6\,\left ({\frac
{\partial ^{2}}{\partial x\partial y}}f(x,y,z,p,q)\right )\left ({
\frac {\partial ^{2}}{\partial x\partial z}}f(x,y,z,p,q)\right ){
\frac {\partial ^{2}}{\partial y\partial
z}}f(x,y,z,p,q)+\]\[+3\,\left ({ \frac {\partial ^{2}}{\partial
y\partial z}}f(x,y,z,p,q)\right )^{2}{ \frac {\partial
^{2}}{\partial {x}^{2}}}f(x,y,z,p,q)-\]\[-3\,\left ({\frac
{\partial ^{2}}{\partial {y}^{2}}}f(x,y,z,p,q)\right )\left
({\frac {
\partial ^{2}}{\partial {x}^{2}}}f(x,y,z,p,q)\right ){\frac {\partial
^{2}}{\partial {z}^{2}}}f(x,y,z,p,q)+\]\[+3\,\left ({\frac
{\partial ^{2}}{
\partial x\partial z}}f(x,y,z,p,q)\right )^{2}{\frac {\partial ^{2}}{
\partial {y}^{2}}}f(x,y,z,p,q)=0
\]
or \begin{equation}\label{yang}
 {\frac {\partial ^{2}}{\partial
p\partial q}}f(x,y,z,p,q)-$$$$-3\left [\begin {array}{ccc} {\frac
{\partial ^{2}}{\partial {x}^{2}}}f( x,y,z,p,q)&{\frac {\partial
^{2}}{\partial x\partial y}}f(x,y,z,p,q)&{ \frac {\partial
^{2}}{\partial x\partial z}}f(x,y,z,p,q)
\\\noalign{\medskip}{\frac {\partial ^{2}}{\partial x\partial y}}f(x,y
,z,p,q)&{\frac {\partial ^{2}}{\partial
{y}^{2}}}f(x,y,z,p,q)&{\frac {
\partial ^{2}}{\partial y\partial z}}f(x,y,z,p,q)\\\noalign{\medskip}{
\frac {\partial ^{2}}{\partial x\partial z}}f(x,y,z,p,q)&{\frac {
\partial ^{2}}{\partial y\partial z}}f(x,y,z,p,q)&{\frac {\partial ^{2
}}{\partial {z}^{2}}}f(x,y,z,p,q)\end {array}\right ]=0.
\end{equation}

    The equation (\ref{yang}) is of Monge-Ampere type and after
    the substitution
\[
f(x,y,z,p,q)=U(p,q)E(x,y,z)
\]
takes the form
\[
{\frac {{\frac {\partial ^{2}}{\partial p\partial q}}U(p,q)}{\left
(U( p,q)\right )^{3}}}-3\,{\frac {{\it Hess}(E(x,y,z))}{E(x,y,z)}}
=0.
\]

    So it has particular solutions defined by the equations
\[
3{\it Hess}(E)-\mu E=0
\]
and
\[
\frac{\partial^2 U}{\partial p \partial q}-\mu U^3=0.
\]

   Both of equations have an important applications.

   The second equation is reduction of the $SU(2)$ Yang-Mills
   equation and the first is a Monge-Ampere type of equation having applications
   in theory of Calabi-Yau manifolds.



\end{document}